\documentclass[onecolumn]{revtex4}
\usepackage{graphicx}
\usepackage{dcolumn}
\usepackage{amsmath}
\usepackage{amssymb}
\usepackage{bm}
\usepackage{amsmath}
\usepackage{amsfonts}
\usepackage{amssymb}

\input epsf

\begin{document}

\title{\Large Solar system constraints on alternative gravity theories}

\author{\bf Sumanta~
Chakraborty \footnote{sumantac.physics@gmail.com} }

\affiliation{Department of Physics\\Rajabajar Science College\\92 A.P.C
Road,~Kolkata-700009,~India}

\author{\bf Soumitra SenGupta \footnote{tpssg@iacs.res.in}}

\affiliation{Department of Theoretical Physics\\Indian Association for the Cultivation of Science\\Kolkata-700032, India}

\date{\today}

\begin{abstract}
The perihelion precession of planetary orbits and
the bending angle of null geodesics are estimated for different
gravity theories in string-inspired models. It is shown that, for
dilaton coupled gravity, the leading order measure in the angle of
bending of light comes purely from vacuum expectation value of
the dilaton field which may be interpreted as an indicator of a
dominant stringy effect over the curvature effect. We arrive at similar results for spherically
symmetric solution in quadratic gravity. We also present the perihelion shift and bending of
light in the Einstein-Maxwell-Gauss-Bonnet theory with special reference to the Casimir effect 
and Damour-Polyakov mechanism. Numerical bounds to different coupling parameters in these models are estimated.
\end{abstract}

\maketitle


\section{Introduction:}

The Perihelion precession of planetary orbits and the bending of light by
a massive star were examined a long time ago in the context of
Einstein's general theory of relativity (\cite{Einstein1}--\cite{Bertotti}) 
in a Schwarzschild metric for the background spacetime. Since then, there have been various
proposals of alternative theories of gravitation beyond Einstein's
gravity. Especially in the context of string theory the effects of
the extra scalar field, namely, the dilaton, lead to a new model
of gravity in the low-energy effective theory in four dimensions (see Gibbons and Hull (\cite{Gibbons1}, \cite{Gibbons2}), Hawking
(\cite{Hawking}), Witten (\cite{Witten1}), Mayers et.al.
(\cite{Mayers}, \cite{Callen}) and Chandrasekhar and Hartle
(\cite{Chandrashekhar}). In addition there are some other higher order gravity theories which admit spherical
symmetry for the spacetime. Among them Einstein-Maxwell-Gauss-Bonnet (henceforth referred to as EMGB) theory
(see \cite{Zumino}, \cite{Lovelock}) and spherically symmetric solution in quadratic gravity (see \cite{Yunes2}, \cite{Deser}) are
of utmost importance.\\

In this paper, we present a general formalism based on the perturbative
method to determine the timelike or null geodesics and estimate
the perihelion precession and light bending for alternative
gravity theories. Among them one very important candidate is
dilaton gravity in the presence of a $U(1)$ gauge field which yields a
charged black hole solution in the low-energy limit of string theory
compactified to four dimensions. Such a solution has been analyzed
in great detail by Garfinkle, Horowitz and Strominger \cite{Garfinkle} (see also
Coleman \cite{Coleman}). We have shown that, while in the case of a pure
Reissner-Nordstr\"{o}m (R-N) solution the bending of light, in the
leading order, is similar to the Schwarzschild scenario, it differs significantly for a charged dilaton coupled solution. For
such a solution the bending angle depends nontrivially on the
vacuum expectation value of the dilaton which is a measure of string
coupling.

We have also presented the calculation of the perihelion shift and bending of light for the EMGB
gravity theory where we have a higher order Gauss-Bonnet term and action for an electromagnetic field.
The natural coupling in this type of theory is the inverse of string constant $\alpha$, 
which we have estimated by using known observational results. For the EMGB theory we have also obtained 
that the shift decreases with charge of the black hole as in the R-N scenario; 
however the expression for the bending angle of light differs significantly and increases with the coupling $\alpha$. 
Recently Yunes and Stein (see \cite{Yunes}) have obtained spherically symmetric solution 
in quadratic gravity in the context of dynamical theories. 
We have used that solution to calculate the perihelion shift and bending of light and obtained nontrivial results. 
Our results further reveal that while for R-N spacetime the perihelion shift decreases with increase in electric charge, 
for a dilaton coupled electromagnetic  background it increases with the electric charge.\\

As an illustration, we first very briefly present the perihelion precession and bending of light in a 
Reissner-Nordstr\"{o}m geometry and then extend our calculations for alternative theories of gravity. 
Our analysis includes charged dilaton black hole, the EMGB gravity theory, and a spherically symmetric solution 
in quadratic gravity. In every case we have tried to estimate a bound on different coupling 
parameters which represent the characteristic scale of the corresponding model beyond Einstein gravity. 
The paper ends with a discussion on our results.

\section{Reissner-Nordstr\"{o}m Black Hole}

\subsubsection{Perihelion Precession}

The metric for a R-N black hole is given by (see \cite{Minser}, \cite{Weinberg})

\begin{equation}\label{1}
ds^{2}=-(1-\frac{2M}{r}+\frac{Q^{2}}{r^{2}})dt^{2}+(1-\frac{2M}{r}+\frac{Q^{2}}{r^{2}})dr^{2}+r^{2}d\Omega^{2}
\end{equation}

where $M$ is the mass of the black hole and $Q$ is the charge. Since the metric does not contain $t$ and $\phi$ explicitly,
the corresponding two conserved quantities are the energy and angular momentum such that

\begin{eqnarray}\label{2}
\left.\begin{array}{c}
p_{t}=-mE \\
p_{\phi}=mL
\end{array}\right\}
\end{eqnarray}

Here $m$ is the mass of the particle, $E$ is the energy per particle mass, $L$ is the angular momentum per particle mass, 
and $p_{t}$ and $p_{\phi}$ are the components of four-momentum. In our chosen unit system, mass has same dimension as energy 
and $E$ is dimensionless. We shall use $E,L$ for massive particles and $E_{0},L_{0}$ for massless photons.

Also, we can choose the motion of the particle in the equatorial plane ($\theta=\frac{\pi}{2}$) (see \cite{Scutz}, \cite{Will}) 
such that $p_{\theta}=0$. The geodesic equation can be obtained by using the energy-momentum relation 
$p_{\mu}p^{\mu}=-m^{2}$ and defining $p^{r}=m\frac{dr}{d\lambda}$ where $\lambda$ is some 
affine parameter connected with the proper time of the particle. Then using the metric coefficients and the energy-momentum 
relation we get an expression for $\frac{dr}{d\lambda}$. To eliminate $\lambda$, 
we introduced $p^{\phi}=m\frac{d\phi}{d\lambda}$, and finally by simple algebraic manipulation with a substitution $r=\frac{1}{u}$ we arrive at,

\begin{equation}\label{15}
\left(u^{'}\right)^{2}=\frac{E^{2}}{L^{2}}-(1-2Mu+Q^{2}u^{2})(u^{2}+\frac{1}{L^{2}})
\end{equation}

Here$'$ denotes derivative with respect to  $\phi$. On further differentiation and keeping terms $\sim u^{2}$, we arrive at

\begin{equation}\label{18}
u^{''}+(1+\varepsilon_{1})u=A+\varepsilon u^{2}
\end{equation}

where we have taken $A=\frac{M}{L^{2}}$, $\varepsilon_{1}=\frac{Q^{2}}{L^{2}}$ and $\varepsilon=3M$. 
We assume that after each revolution the period of $\phi$ changes by a small amount, i.e., $\phi_{period}=2\pi [1+\alpha \varepsilon]$, 
where $\alpha$ is the measure of perihelion precession which depends on $A$, $\varepsilon_{1}$ and $\varepsilon$. 
By using the perturbation method, the solution for $u$ is taken to be $u=A+B~cos(1-\alpha \varepsilon)\phi + \varepsilon u_{1}(\phi)$. 
Then, by keeping terms linear in $\varepsilon$ (since $GM/c^{2}$ as well as the quantity $Q^{2}/L^{2}$ are small), we obtain

\begin{equation}\label{22}
u_{1}^{''}+u_{1}=A^{2}-A\frac{\varepsilon_{1}}{\varepsilon}-\frac{\varepsilon_{1}}{\varepsilon}B~cos(1-\alpha
\varepsilon)\phi-B^{2}~cos^{2}\phi +2AB~cos(1-\alpha
\varepsilon)\phi-2\alpha B~cos(1-\alpha \varepsilon) \phi
\end{equation}

Note that the solution of an equation of the
form $u^{''}+u=\sum_{i}A_{i}~cos\omega _{i}\phi$ will be nonresonant when the $cos(1-\alpha \varepsilon)\phi$ term vanishes
(see \cite{Hooft}), which in turn implies $\alpha=A-\frac{\varepsilon_{1}}{2\varepsilon}$. Hence the perihelion precession would be

\begin{equation}\label{24}
\delta \phi=2 \pi
(A-\frac{\varepsilon_{1}}{2\varepsilon})\varepsilon= 2 \pi
A\varepsilon-\pi \varepsilon_{1}
\end{equation}

\begin{figure}
\begin{center}
\label{fig1}

\includegraphics[height=4in, width=4in]{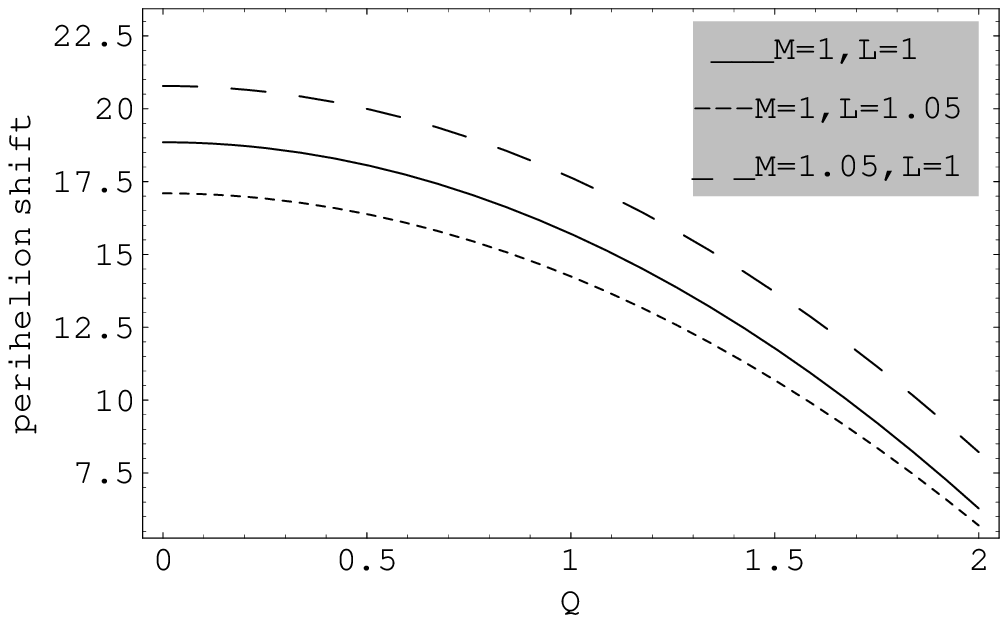}~~~

\vspace{1mm}Figure 1: The figure shows the variation of the perihelion shift with charge.\hspace{1cm}
 \vspace{1mm}

\end{center}
\end{figure}

Thus for the R-N case the shift is given by

\begin{equation}\label{25}
\delta \phi=-\frac{\pi Q^{2}}{L^{2}}+6\pi \frac{M^{2}}{J^{2}}
\end{equation}

Figure 1 depicts how the perihelion shift decreases with an increase in charge of the source.
From Eq. $(\ref{25})$ we see that for a given $\frac{M^{2}}{L^{2}}$ the perihelion shift $\delta \phi$ decreases with $Q$ due
to the negative sign in front. For $Q\rightarrow 0$, the curve in the figure touches the $y$ axis which corresponds to the value of $\delta \phi$
for the Schwarzschild scenario. For example, if we took $M=1,L=1$, then the curve should touch the $y$ axis at $6\pi$, i.e., 
at $\delta \phi =18.9$, which can be observed from the figure.

\subsubsection{Bending of Light}

To estimate the bending of light we can write the null geodesic as

\begin{equation}\label{26}
(\frac{dr}{d\lambda})^{2}=E_{0}^{2}-\left(1-\frac{2M}{r}+\frac{Q^{2}}{r^{2}}\right)\left[\frac{L_{0}^{2}}{r^{2}}\right]
\end{equation}

Simple algebra, as shown in the previous section, leads the above expression to

\begin{equation}\label{30}
\left(\frac{du}{d\phi}\right)^{2}=(u^{'})^{2}=\frac{1}{j_{0}^{2}}-(1-2Mu+Q^{2}u^{2})u^{2}
\end{equation}

where $j_{0}=\frac{L_{0}}{E_{0}}$. Equation ($\ref{30}$) indicates that the term that contains $Q$ 
is of the order of $u^{4}$ and therefore has a very negligible influence on the measure of 
bending of light when the minimum distance of approach is large. 
Note that the term containing $Q^{2}$ also appears in the massive case. 
However there we have a factor of $\frac{1}{L^{2}}$, which makes a nonzero contribution 
from that term. Hence we can conclude that for R-N geometry the bending of light is almost 
identical to that of a Schwarzschild black hole at a sufficiently large distance from the 
source such that terms beyond the leading order, i.e., $u^{3}$ can be ignored.

\section{Alternative Gravity Theories}

\subsection{Dilaton Coupled Electromagnetic field}

Static uncharged black holes in general relativity are described by Schwarzschild solution. 
If the mass of the black hole is large compared to the Planck mass, then this also, to a good approximation, 
describes the uncharged black hole in string theory except regions near a singularity. However, the Einstein-Maxwell 
solution for the string-inspired theory differ widely from the known classical solution due to dilaton coupling.

The dilaton coupling with electromagnetic field tensor $F_{\mu \nu}$ implies that every solution with nonzero 
$F_{\mu \nu} $ will come with a non zero dilaton. Thus the charged black hole solution in general relativity 
(which is the Reissner-Nordstr\"{o}m solution) appears in a new form in string theory due to the presence of the dilaton. 
The effective four-dimensional low-energy Lagrangian obtained from string theory is

$$S=\int d^{4}x \sqrt{-g}[-R+e^{-2\Phi}F^{2}+2(\nabla\Phi)^{2}]$$

where $F^{2}=F_{\mu \nu}F^{\mu \nu}$ is the Maxwell field Lagrangian associated with a $U(1)$ subgroup of 
$E_{8}\times E_{8}$ or $Spin(32)/Z_{2}$. We have set the remaining gauge and antisymmetric tensor field 
$H_{\mu \nu \rho}$ to zero and focus into the presence of $\Phi $, the dilaton field 
(see Garfinkle, Horowitz and Strominger \cite{Garfinkle}, Coleman \cite{Coleman}, Vega and Sanchez \cite{Vega},
 Bekenstein \cite{Bekenstein}, and Witten \cite{Witten2}). Extremizing with respect to the $U(1)$ potential 
$A_{\mu}$, $\Phi$, and $g_{\mu \nu}$, we have the following field equations:

\begin{eqnarray}\label{n1}
\left.\begin{array}{c}
(a) \nabla _{\mu} \left(e^{-2\Phi}F^{\mu \nu} \right)=0\\
(b) \nabla ^{2}\Phi +\frac{1}{2}e^{-2\Phi}F^{2}=0\\
(c) R_{\mu \nu}=2\nabla _{\mu}\Phi \nabla _{\nu}\Phi +2e^{-2\Phi}F_{\mu \lambda}F^{\lambda}_{\nu}-\frac{1}{2}g_{\mu \nu}e^{-2\Phi}F^{2}
\end{array}\right\}
\end{eqnarray}

\subsubsection{Perihelion Precession}

We wish to find a static spherically symmetric solution corresponding to the above field equations 
(\ref{n1}) that are asymptotically flat and have a regular horizon. For this purpose, 
the general metric ansatz can be taken as $ds^{2}=-fdt^{2}+f^{-1}dr^{2}+R^{2}d\Omega ^{2}$, 
where $f$ and $R$ are functions of $r$ only. For a purely magnetic Maxwell field $F=Q~sin\theta d\theta \wedge d\phi$, 
we have $F^{2}=2Q^{2}/R^{4}$ and can deduce that there exist only three independent components for the Ricci tensor, 
namely, $R_{00}$, $R_{11}$, and $R_{22}$. We can also show from Eq. (\ref{n1}c) that $R_{22}=R_{00}$, 
which leads to $(f^{2}R^{2})^{''}=2$, with the prime denoting derivative with respect to the radial coordinate. 
Again from Eq. (\ref{n1}) we find the equation, $R_{00}=-\nabla ^{2}\phi$. 
Using these equations and (\ref{n1}), we can find out the spherically symmetric solution 
(see Garfinkle, Horowitz and Strominger \cite{Garfinkle})

\begin{equation}\label{31}
ds^{2}=-(1-\frac{2M}{r})dt^{2}+\frac{1}{(1-\frac{2M}{r})}dr^{2}+r(r-e^{2\Phi_{0}}\frac{Q^{2}}{M})d\Omega^{2}
\end{equation}

where, $d\Omega^{2}=d\theta^{2}+sin^{2}\theta d\phi^{2}$. Once again due to isometry, 
we have taken our motion in the equatorial plane such that $d\Omega^{2}=d\phi^{2}$. Here $\Phi_{0}$ 
is the asymptotic value of the dilaton, and $Q$ represents the black hole charge. 
Note that this is almost identical to the Schwarzschild metric, with a difference that areas of 
spheres of constant $r$ and $t$ now depend on $Q$. In particular the surface $r=\frac{Q^{2}e^{2\Phi_{0}}}{M}$ 
is singular, and $r=2M$ is the regular event horizon. The evolution of the scalar field $\Phi$ 
may also be derived from the field equations mentioned before Eq. (\ref{31}) and the solution given by

\begin{equation}\label{n2}
e^{-2\Phi}=e^{-2\Phi _{0}}-\frac{Q^{2}}{Mr}
\end{equation}

Hence, as $r \rightarrow \infty$, $\Phi \rightarrow \Phi _{0}$. We can define the dilaton charge as

$$D=\frac{1}{4\pi}\int d^{2}\sigma^{\mu}\nabla_{\mu}\Phi$$

where the integral is over a two sphere at spatial infinity and $\sigma^{\mu}$ is the normal to the two sphere 
at spatial infinity. For a charged black hole we could compute this by using Eq. (\ref{n1}b) and 
the expression for a purely magnetic Maxwell field $F=Q~sin\theta d\theta \wedge d\phi$, which leads to

\begin{equation}\label{n3}
D=-\frac{Q^{2}e^{2\Phi_{0}}}{2M}=-\frac{Q^{2}}{2Me^{-2\phi _{0}}}
\end{equation}

This indicates that the matter particle feels the presence of the dilaton with a coupling modified by 
$e^{-2\phi _{0}}$. Here $D$ depends on the asymptotic value of dilaton field, which is determined once 
$M$ and $Q$ are given and is always negative. Note that the actual dependence on the dilaton field is described by 
$e^{-\Phi /M_{pl}}$. Since we have worked in the unit $M_{pl}\sim 1$, the term is modified to 
$e^{-\Phi}$. As $\Phi \rightarrow \Phi _{0}\sim M_{pl}$, this term is expected to become significant.\\

Following the same procedure as in the previous section, we find for a test particle

\begin{equation}\label{32}
(\frac{dr}{d\lambda})^{2}=-g^{rr}[\epsilon + g^{tt}E^{2} + g^{\phi \phi}L^{2}]
\end{equation}

where $E$ and $L$ are conserved energy and angular momentum respectively. Also, $\epsilon$ is $0$ 
for a massless photon and is $1$ for a massive particle. Defining $A=e^{2\Phi_{0}}\frac{Q^{2}}{M}$ and using

\begin{equation}\label{34}
\frac{d\phi}{d\lambda}=g^{\phi\phi}L=\frac{L}{r(r-A)}
\end{equation}

we arrive at

\begin{equation}\label{36}
(\frac{du}{d\phi})^{2}=\frac{1}{L^{2}}[E^{2}(1-Au)^{2}-\epsilon (1-2Mu)(1-Au)^{2}-u^{2}L^{2}(1-Au)(1-2Mu)]
\end{equation}

For massive particles ($\epsilon=1$), Eq. ($\ref{36}$) becomes

\begin{equation}\label{37}
(\frac{du}{d\phi})^{2}=(\frac{E^{2}-1}{L^{2}})+u[\frac{2(M+A)}{L^{2}}-\frac{2AE^{2}}{L^{2}}]
-u^{2}[1+\frac{A^{2}+4MA}{L^{2}}-\frac{A^{2}E^{2}}{L^{2}}] +u^{3}[2M+A+\frac{2MA^{2}}{L^{2}}]
\end{equation}

Differentiation yields

\begin{equation}\label{38}
u^{''}+(1+\frac{A^{2}+4MA-A^{2}E^{2}}{L^{2}})u=\frac{M+A-AE^{2}}{L^{2}} + \frac{3}{2}u^{2}(2M+A+\frac{2MA^{2}}{L^{2}})
\end{equation}

Note that in the limit $A \rightarrow 0$ the above equation reduces to the orbit equation for 
Schwarzschild solution. Just as in the previous case, here also the quantity $A/L$ is small since 
the quantity $A$ contains mass of the source in inverse power.\\

By comparing with Eq. ($\ref{18}$), the perihelion precession is now estimated as

\begin{equation}\label{39}
\delta \phi = 6\pi \frac{M^{2}}{L^{2}}[1+\frac{Q^{2}}{M^{2}}(1-E^{2})e^{2\Phi_{0}}][1+\frac{Q^{2}}{2M^{2}}e^{2\Phi_{0}}+
\frac{Q^{4}}{M^{2}L^{2}}e^{4\Phi_{0}}] - \frac{\pi Q^{2}}{ML^{2}}e^{2\Phi_{0}}[4M+\frac{Q^{2}}{M}(1-E^{2})e^{2\Phi_{0}}]
\end{equation}

Neglecting terms of the order of $Q^{4}$, we obtain

\begin{equation}\label{40}
\delta\phi = 6\pi \frac{M^{2}}{L^{2}}-12\pi \frac{MD}{L^{2}}\left[\frac{5}{6}-E^{2}\right]
\end{equation}

Again for $D=0$ it reduces to the Schwarzschild case. From Eq. ($\ref{40}$) it is evident that the 
perihelion shift increases with the asymptotic value of dilaton, which is shown explicitly in
Fig. 2.

\begin{figure}

\includegraphics[height=4in, width=4in]{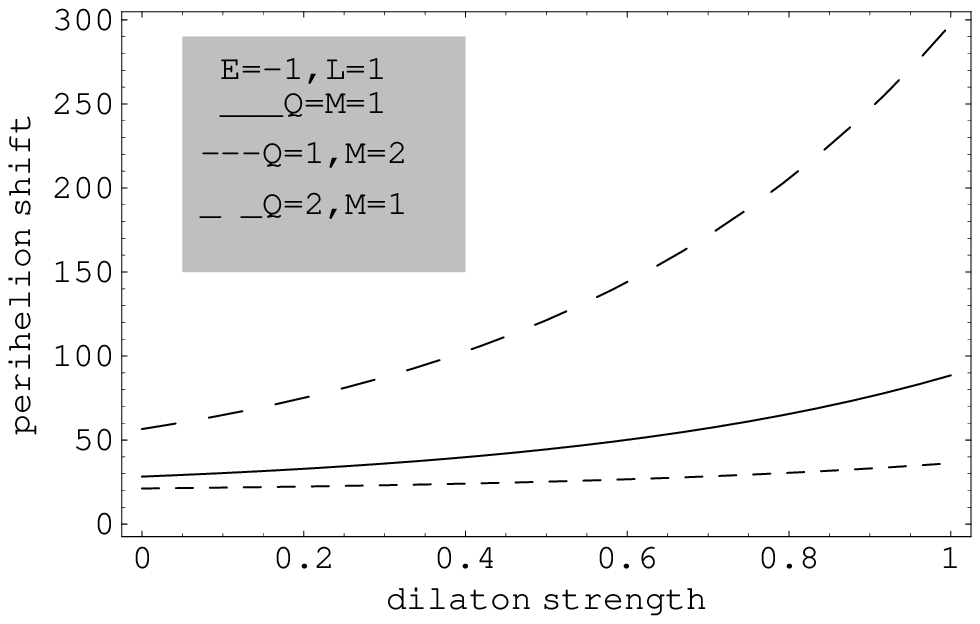}~~~

\vspace{1mm}Figure 2: The figure shows the variation of perihelion
shift with a dilaton field for different choices of $Q$ and $M$.
\hspace{1cm}
 \vspace{1mm}

\end{figure}

It may be observed that the perihelion shift increases very fast with an increase on dilaton strength 
and also increases with an increase in charge of the star or the black hole.

Note that both the R-N solution and the dilaton solution describe a black hole with mass $M$ and charge $Q$ 
only when $Q/M$ is small; otherwise they describe naked singularity. However, we have an important difference 
between these solutions, as in this case there is no analog of the inner horizon which exists for the R-N scenario. 
But, for string theory, the statement that the horizon is singular when $Q^{2}=e^{-2\Phi _{0}}M^{2}$ is actually 
irrelevant, Since the strings do not couple to the metric $g_{\mu \nu}$ but rather to $e^{2\Phi} g_{\mu \nu}$. 
This metric appears in the string $\sigma$ model. In terms of the string metric, the effective Lagrangian becomes (see \cite{Garfinkle})

$$S=\int d^{4}x \sqrt{-g}e^{-2\Phi}\left[-R-4(\nabla \Phi)^{2}+F^{2} \right]$$

Hence the charged black hole metric is obtained as

\begin{equation}\label{n4}
ds^{2}_{string}=-\frac{1-2Me^{\Phi _{0}}/\rho}{1-Q^{2}e^{3\Phi _{0}}/M\rho}d\tau ^{2}+\frac{d\rho ^{2}}{\left(1-2Me^{\Phi _{0}}/\rho \right)\left(1-Q^{2}e^{3\Phi _{0}}/M\rho\right)}+\rho ^{2}d\Omega
\end{equation}

This metric is identical to the metric given in Eq. $(\ref{31})$ where we have just rescaled the metric by some 
conformal factor which is finite everywhere outside and on the horizon. If we try to calculate the perihelion precession 
by using this metric following the same method, we arrive at the same result presented by Eq. $(\ref{40})$.

From the available experimental data on the perihelion precession of Mercury, the error bar is estimated as 
$3$ arcsec/century. So we can estimate a bound for the dilaton charge, which is $0 \leq D \leq 1.29979\times 10^{-13}$. 
This clearly explains why the stringy signature can still not be determined within the precision 
of the present day astrophysical experiments.

\subsubsection{Bending of Light}

\begin{figure}

\includegraphics[height=4in, width=4in]{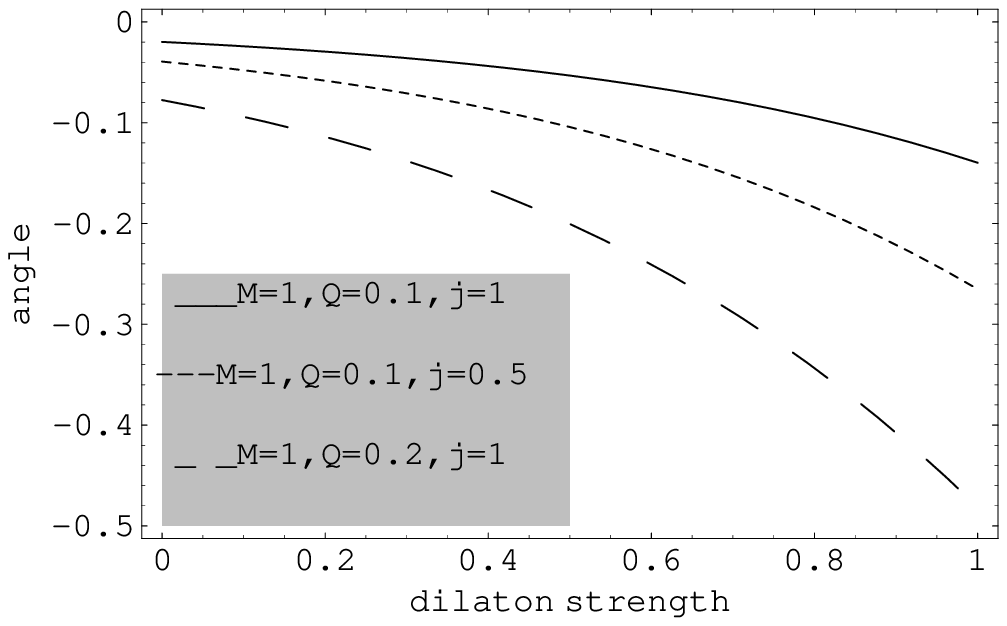}~~~

\vspace{1mm}Figure 3: The figure shows the variation of bending of
light with a dilaton strength for different choices of mass
and charge.\hspace{1cm}
 \vspace{1mm}

\end{figure}

For the null geodesic the orbit equation yields

\begin{equation}\label{41}
(\frac{du}{d\phi})^{2}=\frac{E_{0}^{2}}{L_{0}^{2}}(1-2Au+A^{2}u^{2})-u^{2}+(A+2M)u^{3}-2MAu^{4}
\end{equation}

Introducing $j_{0}=L_{0}/E_{0}$, i.e., angular momentum per unit energy or, equivalently, angular momentum per particle mass, we obtain

\begin{equation}\label{42}
\frac{d\phi}{du}=\frac{1}{\sqrt{\frac{1}{j_{0}^{2}}-\frac{2Au}{j_{0}^{2}}+u^{2}(\frac{A^{2}}{j_{0}^{2}}-1)+(A+2M)u^{3}}}
\end{equation}

By neglecting terms of order $u^{3}$, the above differential equation can be integrated such that

\begin{equation}\label{43}
\phi -\phi_{0}=\int_{u_{0}}^{u}\frac{du}{\sqrt{\frac{1}{j_{0}^{2}}-c_{1}u-c_{2}u^{2}} }
\end{equation}

where $c_{1}=\frac{2A}{j_{0}^{2}}$ and $c_{2}=(1-\frac{A^{2}}{j_{0}^{2}})$. The lower limit $u_{0}$ 
represents the position of the photon when it does not feel the presence of the field, i.e., at $r\rightarrow\infty$. 
The upper limit yields the minimum distance of approach, which in this case is the solution of the equation $u^{'}=0$. 
Using these we determine the bending of light as

\begin{equation}\label{43a}
\Delta \phi =2F(A/j_{0})-\pi
\end{equation}

where $F(x)=cos^{-1}(x)/\sqrt{1-x^{2}}$. Note that in the limit $A\rightarrow 0$ this vanishes, 
which is consistent, since in Eq. ($\ref{42}$) we have eliminated the $u^{3}$ term which appears 
in the Schwarzschild case. As we have neglected the dominant term in the Schwarzschild case ($\sim u^{3}$), 
we should not expect the Schwarzschild result to be retrieved. 
Thus, the above bending is dependent crucially on the dilaton asymptotic value and comes from a more dominant term 
($\sim u^{2}$) than that in the Schwarzschild scenario. So this bending can be interpreted as a signature 
of a string-inspired dilaton spacetime. Here $A=e^{2\Phi_{0}}\frac{Q^{2}}{M}=-2D$ [by Eq. (\ref{n3}] 
where $D$ is the dilaton charge. Using this we finally obtain

\begin{equation}\label{44}
\Delta \phi =2F(-2D/j_{0})-\pi
\end{equation}

with $F(x)$ defined after Eq. (\ref{43a}). However, the above solution is valid for $j_{0}^{2}>4D^{2}$. 
For the other choice the bending of light is given by

\begin{equation}\label{45}
\Delta \phi =2F(-2D/j_{0})-\pi
\end{equation}

where $F(x)=cosh^{-1}(x)/\sqrt{x^{2}-1}$. 
This also vanishes in the limit $D \rightarrow 0$ and hence is consistent with the above analysis. 
The first solution is applicable when the $L_{0}/E_{0}$ of the particle is much greater than the dilaton strength, 
and the second condition is just the reverse of the first condition. 
Following the classic test of general relativity by Cassini spacecraft \cite{Bertotti}, 
we could put bounds on the dilaton charge $D$. The quantity $\gamma$ could be calculated as 
$\gamma -1 =\Delta \phi _{dilaton}-\Delta \phi _{GR}$. Then, following the constraint put on this 
$\gamma -1$ by Cassini \cite{Bertotti}, we get, $D\leq \bigcirc (10^{-7})$. 
This bound on $D$ is distinctly several orders larger than the bound derived from the perihelion precession test. 
However one should also need to consider Damour-Polyakov mechanism, which generates a screening effect 
and may keep the estimate within experimental bounds (\cite{Damour1}, \cite{Damour2}). 

String-loop corrections of low-energy matter couplings of the dilaton provide a mechanism 
to fix the vacuum expectation value of the massless dilaton, compatible with existing experimental data. 
By using some assumptions regarding the universal nature of the dilaton coupling function, 
it is possible to show that the cosmological evolution of a graviton-dilaton-matter system drives 
the dilaton to a value when it decouples from matter, called the least coupling principle \cite{Damour1}. 
The origin of this evolution is explained as follows: Masses of different particles depend on the dilaton $\phi$, 
while the source for the dilaton mass is the gradient of $\phi$. 
It is therefore not surprising to have a fixed point where the gradient of $\phi$ is zero leading to a massless dilaton. 
This mechanism is similar to the generic attractor mechanism in scalar tensor theories discussed in Ref. \cite{Damour3}. 
Other types of couplings of a long-range scalar field $\phi$ atre given as a universal 
multiplicative coupling of $\phi$ to all the other fields, 
where $L_{tot}=B(\Phi)L_{0}(g_{\mu \nu}, \nabla \Phi ,A_{\mu}, \psi ,\cdots)$ with $B(\Phi)$ 
admitting a universal extremum. 

The universal coupling function $B(\Phi)$ can be expanded in powers of $e^{2\Phi}$ in the small coupling limit, 
i.e., $g_{s}\rightarrow 0$ or, equivalently, $\Phi \rightarrow -\infty$ such that \cite{Damour2}

\begin{equation}\label{dp}
 B(\Phi)=e^{-2\Phi}+c_{0}+c_{1}e^{2\Phi}+c_{2}e^{4\Phi}
\end{equation}

In the low-energy limit of the underlying string theory, which is the one we are discussing here, or equivalently 
the coupling constant, $g_{s}\rightarrow 0$  leads to the first term in the above equation to be the dominant one validating our 
results. In this limit we may ignore the backreaction of the other terms in the expansion of $B(\Phi)$ 
on the metric and can use the form of the metric discussed in ($\ref{n4}$).

In the low-energy limit the original Damour-Polyakov action written in the ``string frame'' can be reduced to that 
in the ``Einstein'' frame with a conformal transformation such that the action in the Einstein frame is identical to that of the 
effective low-energy Lagrangian used in this section \cite{Damour1}. This justifies the use of the Damour-Polyakov mechanism 
to lower the bounds on the dilaton charge.

We summarize in qualitative terms the basic reasons why a massless dilaton 
is rendered nearly invisible during its cosmological evolution: (i) During the radiation era, 
when the Universe passes through a temperature $T\sim m_{A}$, 
the $A$-type particles and antiparticles become nonrelativistic before annihilating 
and hence disappearing from the cosmic soup; this provides a source term for the dilaton, 
which attracts $\phi$ toward a minimum $\phi _{m}^{A}$ of $m_{A}(\phi)$. 
According to Ref. \cite{Damour1} this attraction mechanism is efficient. 
(ii) In the subsequent matter era, $\phi$ will be continuously attracted toward a minimum 
of the mass function $m_{m}(\phi)$ corresponding to the matter dominating the Universe. 
(Under the same universality condition this minimum will be again $\phi _{m}$.) 
Thus, by, following this Damour-Polyakov mechanism it is possible to reduce the dilaton charge 
beyond the experimentally observable value by the screening effect 
(\cite{Damour1}, \cite{Damour2} and \cite{Damour3}). Hence the effective bound on dilaton charge is 
lower than estimated bound following the mechanism illustrated above.

We, however, reiterate that the angle for the bending of light increases in
magnitude with the dilaton strength (see Fig. 3) and also with the
electric charge.

\subsection{Einstein-Maxwell-Gauss-Bonnet gravity}

Theories with an extra spatial dimension have been an active area of interest ever since the original work of 
Kaluza and Klein and the advent of string theory, which predicts the inevitable presence of an extra spatial dimension. 
Among many alternatives, the brane world scenario is considered as a strong candidate within some theoretical basis 
in some underlying string theory. Usually, the effect of string theory on classical gravitational physics 
(see \cite{Greens}, \cite{Davies}) is investigated by means of a low-energy effective action, which in addition to 
the Einstein-Hilbert action contains squares and higher powers of curvature term. However, 
the field equations become fourth order and bring in ghosts (see \cite{Zumino}). In this context, 
Lovelock (see \cite{Lovelock}) showed that, if the higher curvature terms appear in a particular combination, 
the field equation become second order and consequently the ghosts disappear.

In EMGB gravity, the action in five-dimensional spacetime ($M,g_{\mu \nu}$) can be written as

\begin{equation}\label{46}
S=\frac{1}{2}\int _{M} d^{5}x \sqrt{-g} \left[R+\alpha L_{GB}+L_{matter} \right],
\end{equation}

where $L_{GB}=R_{\alpha \beta \gamma \delta}R^{\alpha \beta \gamma \delta}-4R_{\mu \nu}R^{\mu \nu}+R^{2}$ 
is the GB Lagrangian and $L_{matter}=F^{\mu \nu}F_{\mu \nu}$ is the Lagrangian for the electromagnetic field. 
Here $\alpha$ is the coupling constant of the GB term having dimension (length)$^{2}$. 
As $\alpha$ is regarded as inverse string tension, so $\alpha \geq 0$.

For the gravitational and electromagnetic field equations obtained by varying the above action with respect to 
$g_{\mu \nu}$, and $A_{\mu}$, we obtain (see \cite{Chakraborty})

\begin{eqnarray}\label{47}
\left. \begin{array}{c}
G_{\mu \nu}-\alpha H_{\mu \nu}=T_{\mu \nu}\\
\bigtriangledown _{\mu}F^{\mu}_{\nu}=0\\
H_{\mu \nu}=2\left[RR_{\mu \nu}-2R_{\mu \lambda}R^{\lambda}_{\mu}-2R^{\gamma \delta}R_{\mu \gamma \nu \delta}+R^{\alpha \beta \gamma}_{\mu}R_{\nu \alpha \beta \gamma} \right]-\frac{1}{2}g_{\mu \nu}L_{GB}
\end{array}\right \}
\end{eqnarray}

where $T_{\mu \nu}=2F^{\lambda}_{\mu}F_{\lambda \nu}-\frac{1}{2}F_{\lambda \sigma}F^{\lambda \sigma}g_{\mu \nu}$ 
is the electromagnetic field tensor.

A spherically symmetric solution to the above action was obtained by Dehghani (see \cite{Dehghani}), 
and the corresponding line element is given by

\begin{equation}\label{48}
ds^{2}=-g(r)dt^{2}+\frac{dr^{2}}{g(r)}+r^{2}d\Omega _{3}^{2},
\end{equation}

where the metric coefficient is

\begin{equation}\label{49}
g(r)=K+\frac{r^{2}}{4\alpha}\left[1\pm \sqrt{1+\frac{8\alpha \left(m+2\alpha \mid K \mid \right) }{r^{4}} -\frac{8\alpha q^{2}}{3r^{6}}} \right]
\end{equation}

It should be noted that, under asymptotic expansion (this is identical to small $\alpha$ expansion), 
the function $g(r)$ can be written as

\begin{equation}\label{49a}
g_{\infty}(r)=K+\frac{r^{2}}{4\alpha}(1\pm 1) \pm \frac{m+2 \mid K \mid \alpha}{r^{2}} \mp \frac{q^{2}}{3r^{4}}
\end{equation}

Here $K$ is the curvature, $m+2\alpha \mid K\mid$ is the geometrical mass, and $d\Omega _{3}^{2}$ is the metric of a 
3D hypersurface such that

\begin{equation}\label{50}
d\Omega _{3}^{2}=d\theta _{1}^{2}+sin^{2}\theta _{1}\left( d\theta _{2}^{2}+sin^{2}\theta _{2}d\theta _{3}^{2}\right)
\end{equation}

The range is given by $\theta _{1},\theta _{2}:[0,\pi]$. Note that the Gauss-Bonnet term decreases the mass of the 
spacetime for negative $\alpha$ and increases for positive $\alpha$. It is worthwhile to mention 
that this happens only in five-dimensional spacetime \cite{Dehghani}.

We assume that there is a constant charge $q$ at $r=0$ and the vector potential $A_{\mu}=\Phi (r)\delta_{\mu}^{0}$ 
such that $\Phi (r)=-\frac{q}{2r^{2}}$ in five-dimensional spacetime. For four-dimensional effective theory, 
this reduces to the usual form $\Phi (r)=-q/r$.

In this metric the metric function $g(r)$ will be real for $r \geq r_{0}$, 
where $r_{0}^{2}$ is the largest real solution of the cubic equation

\begin{equation}\label{51}
3z^{3}+24\alpha \left(m+2\alpha \mid K \mid \right)z-8\alpha q^{2}=0
\end{equation}

By a transformation of the radial coordinate we can show that $r=r_{0}$ is an essential singularity of the spacetime.

We shall consider the negative sign in front of the square root in Eq. (\ref{49}) since the positive sign does 
not lead to an asymptotically flat solution \cite{Dehghani}. One may note that for $K=0$ the function 
$g(r)$ tends to $0$ as $r$ tends to infinity and is not an acceptable solution. For $K=1$ the function 
$g(r)$ tends to $1$ as $r$ tends to infinity and thus we have an asymptotically flat solution. 
Under this condition Eq. (\ref{49a}) reduces to

\begin{equation}\label{51a}
g(r)=1-\frac{m+2\alpha}{r^{2}}+\frac{q^{2}}{3r^{4}}
\end{equation} 

Thus the presence of an additional dimension leads to a Reissner-Nordstr\"om black hole with mass parameter $m+2\alpha$ \cite{Dehghani}.

\subsubsection{Perihelion Precession}

In this scenario, we shall take the following choice, which we have used in the previous two sections 
by exploiting the spherical symmetry such that $\theta _{1}=\frac{\pi}{2}$, $\theta _{2}=\frac{\pi}{2}$, and $\theta _{3}=\phi$.
Then the equation of motion is given by

\begin{equation}\label{52}
\left(\frac{dr}{d\phi}\right)^{2}=\frac{r^{4}}{L^{2}}\left[E^{2}-g(r)\left(\epsilon +\frac{L^{2}}{r^{2}}\right) \right]
\end{equation}

Since we are interested in the trajectory of massive particles, we choose $\epsilon =1$, 
defining the variable $r=\frac{1}{u}$, and the above equation of motion reduces to

\begin{equation}\label{53}
\left(\frac{du}{d\phi}\right)^{2}=\frac{1}{L^{2}}\left[E^{2}-g(u)\left(1 +L^{2}u^{2}\right) \right]
\end{equation}

where we have

\begin{equation}\label{54}
g(u)=1-\frac{1}{4\alpha u^{2}}\left[1-\sqrt{1+8\alpha \left(m+2\alpha\right)u^{4} -\frac{8\alpha q^{2}}{3}u^{6}} \right]
\end{equation}

Differentiating the above expression with respect to the angular coordinate, we readily obtain the following differential equation

\begin{equation}\label{55}
\frac{d^{2}u}{d\phi ^{2}}=-ug(u)-\frac{1}{2L^{2}}\frac{dg}{du}\left(1+L^{2}u^{2}\right)
\end{equation}

This equation can be simplified in the large $r$ or small $u$ limit as

\begin{equation}\label{56}
\frac{d^{2}u}{d\phi ^{2}}+\left(1-\frac{m+2\alpha}{L^{2}} \right)u=\left[2\left(m+2\alpha \right)-\frac{2q^{2}}{3L^{2}}\right]u^{3}
\end{equation}

Now, following the same procedure, i.e., assuming a periodic solution as in the R-N scenario and rewriting the differential equation as

\begin{equation}\label{56a}
\frac{d^{2}u}{d\phi ^{2}}+\left(1+\epsilon _{1}\right)u=\epsilon u^{3}
\end{equation}

where we have taken $\epsilon _{1}=-\frac{m+2\alpha}{L^{2}}$ and $\epsilon =2(m+2\alpha) -\frac{2q^{2}}{3L^{2}}$,
we finally obtain the following expression for the precession angle:

$$\delta \phi = \frac{\pi \left(m+2\alpha\right)}{L^{2}}+\frac{3}{4}\frac{\pi}{L^{2}}\left[2(m+2\alpha)-\frac{2}{3}\frac{q^{2}}{L^{2}} \right]$$

This can be rewritten by algebraic simplification as

\begin{equation}\label{58}
\delta \phi =\frac{5}{2}\pi \left(\frac{m+2\alpha}{L^{2}}\right)-\frac{1}{2}\pi \left(\frac{q^{2}}{L^{4}}\right)
\end{equation}

The solution though asymptotically flat, does not reduce to the Schwarzschild solution in the limit 
$\alpha \rightarrow 0$, since the action modified by quadratic terms is not the usual Einstein-Hilbert 
action even if $\alpha =0$ (see \cite{Lovelock}). Note that the perihelion shift varies linearly with 
the coupling parameter $\alpha$. As coupling increases, the shift also increases. However, in this 
scenario also the shift decreases with the charge as in the R-N scenario. In this case, we can estimate 
the parameter $\alpha$ by using the estimate of 43 arcsec/century. Then this poses a limit on $\alpha$
which is $0 \leq \alpha \leq 1.2725 \times 10^{-7}$. Hence, this effect is also very small in respect
to the present experimental resolution.

\subsubsection{Bending of Light}

Now we shall discuss the null geodesic solution to the above metric.
This can be obtained easily by substituting $\epsilon =0$ in Eq. $(\ref{52})$, which leads to

\begin{equation}\label{59}
\left(\frac{du}{d \phi} \right)^{2}=\frac{E_{0}^{2}}{L_{0}^{2}}-u^{2}\left[1-(m+2\alpha)u^{2}\right]
\end{equation}

Here we have kept terms up to $u^{4}$, and $E_{0}$ and $L_{0}$ are the energy and angular momentum of the photon, respectively. 
Then from this the bending angle can be determined as

\begin{equation}\label{60}
\Delta \phi = \int _{u_{0}}^{u} \frac{du}{\sqrt{\frac{1}{j_{0}^{2}}-u^{2}\left[1-(m+2\alpha)u^{2}\right]}}
\end{equation}

where $j_{0}=\frac{L_{0}}{E_{0}}$, $u_{0}$ represents the asymptotic value from where the photon came, 
and the upper limit is the minimum distance of approach. Then following the procedure presented in a work 
by Chakraborty and Chakraborty (see \cite{Chakraborty1}), we finally obtain the bending angle as

\begin{equation}\label{61}
\Delta \phi = - \frac{3\pi\left(m+2\alpha\right)}{4j_{0}^{2}}
\end{equation}

Hence, we find that bending increases in magnitude with the coupling $\alpha$, which is expected, 
since the increase of coupling implies the increase in the strength of the field.

\subsection{Spherically symmetric solution in quadratic gravity}

Testing strong-field features of general relativity (GR) is very important for astrophysical signatures, 
as they could indicate departure from general relativity with deep implications in fundamental theory. 
Thus in order to test such features we need deviations from Schwarzschild or Kerr and look for non-GR solutions. 
These non-GR solutions are known through numerical studies, where one choose an alternative theory, 
constructs the field equations, and by postulating a metric ansatz tries to address various astrophysical 
signatures. The differential equations satisfied by these functions are solved and studied numerically (see \cite{Kanti}, \cite{Pani}). 
We can solve those differential equations by approximation methods; for example, we can expand in 
terms of the coupling constants of the theory. This small-coupling approximation (see \cite{Yunes3}) treats 
the alternative theory as an effective and approximate model that allows for small perturbation about GR. 
In this section we consider a class of alternative theories of gravity in four dimensions defined by modifying 
the Einstein-Hilbert action through all possible quadratic, algebraic curvature scalars, 
multiplied by constants or nonconstant couplings as (see \cite{Yunes}),\\

$S=\int d^{4}x \sqrt{-g}[\kappa R+\alpha _{1}f_{1}(\upsilon)R^{2}+\alpha _{2}f_{2}(\upsilon)R_{ab}R^{ab}+\alpha _{3}f_{3}(\upsilon)R_{abcd}R^{abcd}$

\begin{equation}\label{62}
+\alpha _{4}f_{4}(\upsilon)R_{abcd}^{*}R^{abcd}-\frac{\beta}{2}\left(\nabla _{a}\upsilon \nabla ^{a}\upsilon +2V(\upsilon)\right)+L_{matter}]
\end{equation}

where $g$ is the determinant of the metric $g_{ab}$; $R$, $R_{ab}$, $R_{abcd}$, $R_{abcd}^{*}$ are the Ricci scalar and tensor 
and the Riemann tensor and its dual (see \cite{Yunes2}), respectively; $L_{matter}$ is the Lagrangian density for other matter; 
$\upsilon$ is a scalar field; $(\alpha _{i},\beta)$ are coupling constants; and $\kappa=(16\pi G)^{-1}$. 
All other quadratic curvature terms are linearly dependent on these terms. Theories of this type are motivated from 
low-energy expansion of string theory (see \cite{Deser}, \cite{Green}).

Varying Eq. $(\ref{62})$ with respect to the metric and setting $f_{i}(\upsilon)=1$, we find the modified field equations

\begin{equation}\label{63}
\kappa G_{ab}+\alpha _{1}H_{ab}+ \alpha _{2}I_{ab}+ \alpha _{3}J_{ab}=\frac{1}{2}T_{ab}^{matter}
\end{equation}

where $T_{ab}^{matter}$ is the stress energy of matter and

\begin{eqnarray}\label{64}
\left. \begin{array}{c}
(a) H_{ab}=2R_{ab}R-\frac{1}{2}g_{ab}R^{2}- 2 \nabla _{ab}R+ 2g_{ab}\square R\\
(b) I_{ab}=\square R_{ab}+2R_{abcd}R^{cd}-\frac{1}{2}g_{ab}R_{cd}R^{cd}+\frac{1}{2}g_{ab}\square R -\nabla _{ab}R,\\
(c) J_{ab}=8R^{cd}R_{acbd}-2g_{ab}R^{cd}R_{cd}+4\square R_{ab}-2RR_{ab}+\frac{1}{2}g_{ab}R^{2}-2\nabla _{ab}R
\end{array}\right \}
\end{eqnarray}

with $\nabla _{a}$, $\nabla _{ab}(=\nabla _{a}\nabla _{b})$, and $\square (= \nabla _{a}\nabla ^{a})$ are the first- and second-order 
covariant derivative and the D'Alembertian.
The scalar field equation can be given by

\begin{equation}\label{64a}
\beta \square \upsilon -\beta \frac{dV}{d\upsilon}=-\alpha _{1}R^{2}-\alpha _{2}R_{ab}R^{ab}-\alpha _{3}R_{abcd}R^{abcd}- \alpha _{4}R_{abcd}^{*}R^{abcd}
\end{equation}\\

The spherically symmetric solution to the above field equations, imposing dynamical arguments, 
could be written using the metric ansatz as (see \cite{Yunes})

\begin{equation}\label{65}
ds^{2}=-f_{0}\left[1+\epsilon h_{0}(r)\right]dt^{2}+ f_{0}^{-1}\left[1+\epsilon k_{0}(r)\right]dr^{2}+r^{2}d\Omega ^{2}
\end{equation}

and $\upsilon = \upsilon _{0}+\epsilon \upsilon _{0}$, where $f_{0}=1-2M_{0}/r$, with $M_{0}$ the bare mass and $d\Omega _{2}$ 
the line element on the two sphere. The free functions $(h_{0},k_{0})$ are small deformations about the Schwarzschild metric.

The scalar field equation can be solved to yield

\begin{equation}\label{66}
\upsilon _{0}=\frac{\alpha _{3}}{\beta}\frac{2}{M_{0}r}\left(1+\frac{M_{0}}{r}+\frac{4M_{0}^{2}}{3r^{2}} \right)
\end{equation}

We can use this scalar field solution to solve modified field equations to an order linear in $\epsilon$. 
Requiring the metric to be asymptotically flat and regular at $r=2M_{0}$, we find the unique solution 
$h_{0}=F\left(1+\tilde{h_{0}}\right)$ and $K_{0}=-F\left(1+\tilde{h_{0}}\right)$, where $F=-(49/40)\zeta (M_{0}/r)$ and

$$\tilde{h_{0}}=\frac{2M_{0}}{r}+\frac{548}{147}\frac{M_{0}^{2}}{r^{2}}+\frac{8}{21}\frac{M_{0}^{3}}{r^{3}}-
\frac{416}{147}\frac{M_{0}^{4}}{r^{4}}-\frac{1600}{147}\frac{M_{0}^{5}}{r^{5}}$$

\begin{equation}\label{67}
\tilde{k_{0}}=\frac{58}{49}\frac{M_{0}}{r}+\frac{76}{49}\frac{M_{0}^{2}}{r^{2}}-\frac{232}{21}\frac{M_{0}^{3}}{r^{3}}-
\frac{3488}{147}\frac{M_{0}^{4}}{r^{4}}-\frac{7360}{147}\frac{M_{0}^{5}}{r^{5}}
\end{equation}

Here we have defined the dimensionless coupling function $\zeta=\frac{\alpha _{3}^{2}}{\beta \kappa M_{0}^{4}}$, 
which is of the order of $\epsilon$. Such a solution is most general for all dynamical, algebraic, 
quadratic gravity theories, in spherical symmetry. We can define the physical mass $M=M_{0}\left[1+(49/80)\zeta\right]$ 
such that the modified metric components become $g_{tt}=-f(1+h)$ and $g_{rr}=f^{-1}(1+k)$
where $h=\zeta /(3f)(M/r)^{3}\tilde{h}$ and $k=-(\zeta / f)(M/r)^{2}\tilde{k}$. Here,

\begin{equation}\label{68}
\tilde{h}=1+\frac{26M}{r}+\frac{66}{5}\frac{M^{2}}{r^{2}}+\frac{96}{5}\frac{M^{3}}{r^{3}}-\frac{80M^{4}}{r^{4}}
\end{equation}

\begin{equation}\label{69}
\tilde{k}=1+\frac{M}{r}+\frac{52}{3}\frac{M^{2}}{r^{2}}+\frac{2M^{3}}{r^{3}}+ \frac{16M^{4}}{5r^{4}}- \frac{368}{3}\frac{M^{5}}{r^{5}}
\end{equation}

where $f=1-2M/r$. Note from the above expression for metric element that physical observables are related to 
renormalized mass $M$, not bare mass $M_{0}$.

\subsubsection{Perihelion Precession}

In this metric ansatz we shall use the choice $\theta =\pi /2$ exploiting the spherical symmetry of the solution. 
Then the orbit equation for a massive particle would be given in the $r-\phi$ plane to yield

\begin{equation}\label{70}
\left(\frac{dr}{d\phi} \right)^{2}=-\frac{r^{4}}{L^{2}}\frac{1}{g_{rr}}\left[1+\frac{E^{2}}{g_{tt}}+\frac{L^{2}}{r^{2}} \right]
\end{equation}

Introducing the new variable $r=1/u$ we rewrite the orbit equation, using the metric co-efficient as,

\begin{equation}\label{71}
\left(\frac{du}{d\phi}\right)^{2}=- \frac{1+L^{2}u^{2}}{L^{2}}\frac{f(u)}{1+k(u)}+ \frac{E^{2}}{L^{2}}\frac{1}{\left[\left(1+k(u)\right)\left(1+h(u)\right)\right]}
\end{equation}

By substituting for $h(u)$ and $k(u)$ and using the fact that $u$ is a small quantity, the above orbit equation is simplified to

\begin{equation}\label{72}
\left(\frac{du}{d\phi}\right)^{2}=- \frac{1+L^{2}u^{2}}{L^{2}}\left(1-2Mu\right)
\left[1+\zeta M^{2}u^{2}+3\zeta M^{3}u^{3}\right]+ \frac{E^{2}}{L^{2}}\left[1+\zeta M^{2}u^{2}+\frac{8}{3}\zeta M^{3}u^{3}\right]
\end{equation}

This equation can be rearranged to yield

\begin{equation}\label{73}
\left(\frac{du}{d\phi}\right)^{2}=\frac{E^{2}-1}{L^{2}}+ \frac{2M}{L^{2}}u + \left[-1- \frac{\zeta M^{2}}{L^{2}}(1-E^{2})\right]u^{2} + \left[2M- \frac{\zeta M^{3}}{L^{2}}(1-8/3 E^{2})\right]u^{3}
\end{equation}

Differentiating this equation again, we finally obtain,

\begin{equation}\label{74}
\frac{d^{2}u}{d\phi ^{2}}+\left(1+ \zeta \frac{M^{2}}{L^{2}}(1-E^{2})\right)u=\frac{M}{L^{2}}+\frac{3}{2}u^{2}
\left[2M- \frac{M^{3}\zeta}{L^{2}}\left(1-8/3 E^{2}\right) \right]
\end{equation}

Now taking a trial solution of the form $u=A+B~cos(1-\alpha \varepsilon)\phi + \varepsilon u_{1}(\phi)$ and keeping terms linear 
in $\varepsilon$, we arrive at the following expression for
perihelion precession:

\begin{equation}\label{75}
\delta \phi = 6\pi \frac{M^{2}}{L^{2}}- \pi \zeta \frac{M^{2}}{L^{2}}\left(1-E^{2}\right)-3\pi \zeta \frac{M^{4}}{L^{4}}\left(1- 8/3 E^{2}\right)
\end{equation}

As a check note that as $\zeta =0$ we have $M=M_{0}$ and hence the above expression for the perihelion shift 
reduces to the Schwarzschild result. Rewriting the above expression in terms of the bare mass, we finally obtain

\begin{equation}\label{76}
\delta \phi = 6\pi \frac{M_{0}^{2}}{L^{2}}+6\pi \left(\frac{49}{40}\zeta \right)\frac{M_{0}^{2}}{L^{2}} - 
\pi \zeta \left(1-E^{2}\right)\frac{M_{0}^{2}}{L^{2}}-3\pi \zeta \left(1-\frac{8}{3}E^{2} \right)\frac{M_{0}^{4}}{L^{4}}
\end{equation}

From this equation we see that the second term in the right-hand side expression increases with $\zeta$, 
while the other two terms decrease. However if $E$ is greater than 1 then all terms are positive. 
So the perihelion shift depends on the energy of the particle. Also since we have kept terms linear in 
$\zeta$ (the metric itself is valid for linear terms in $\zeta$; see \cite{Yunes}), the shift varies linearly, 
but whether it will increase or decrease depends on the energy value.

In this case, we can estimate the parameter $\zeta$ using the fact that perihelion precession 
measurements for Mercury show an error bar of 3 arcsec/century. Then using the above expression for the 
perihelion precession we can limit $\zeta$ to a small value of $0 \leq \zeta \leq 6.59 \times 10^{-4}$. 
So the effect is not as small as in the above two cases, though not significant enough to be within the present day experimental observation.

\subsubsection{Bending of Light}

The null geodesic in this case is given by

\begin{equation}\label{77}
\left(\frac{du}{d\phi} \right)^{2}=-u^{2}\left(1-2Mu\right) \left(1+\zeta M^{2}u^{2} \right)+\frac{E_{0}^{2}}{L_{0}^{2}}
\end{equation}

where we have kept terms up to the order of $u^{2}$, which is the leading order contribution.
Also $E_{0}$ and $L_{0}$ are the energy and angular momentum, respectively, of the photon under consideration.
This equation can be integrated to obtain the bending angle as

\begin{equation}\label{78}
\Delta \phi = \int _{u_{0}}^{u}\frac{du}{\sqrt{\frac{1}{j_{0}^{2}}- \left(1-\zeta \frac{M^{2}}{j_{0}^{2}}\right)}}
\end{equation}

Here $u_{0}$ represents the asymptotic value of the inverse radius from where the photon approaches.
Since the solution is asymptotically flat this value is $0$ to be precise.
However the upper limit implies the minimum distance i.e. the solution of the equation
$u^{'}=0$. Then the above equation can be integrated to yield

\begin{equation}\label{79}
\Delta \phi = \pi - \frac{\pi}{\sqrt{1-\zeta \frac{M^{2}}{j_{0}^{2}}}}
\end{equation}

where $j_{0}=L_{0}/E_{0}$. In this case the bending depends on the leading order term $\sim u^{2}$ 
and we have kept terms just up to that order. So in the limit of $\zeta \rightarrow 0$ the bending angle of 
light vanishes. This shows that the parameter $\zeta$ provides a longer range effect on bending of light 
than the Schwarzschild mass does. So we again have an instance where the bending angle is 
coming from a purely string theory effect in a modified Schwarzschild solution in the presence of quadratic gravity.

\section{Conclusions}
In this paper, we have considered various alternative theories of gravity which have their roots 
in some underlying string theory. Considering three such variants, we have explored the influence of various 
modifications beyond Einstein gravity on the perihelion shift and lensing. From the present available data 
for these phenomena, we have established a bound on the characteristic parameter of each model. 
For a dilaton coupled charged solution, our work brings out the possibility of a stringy signature 
and also shows a remarkable feature that the dilaton vacuum expectation value, which decides the magnitude of the 
bending angle (though suppressed by the scale of the theory), has its presence at one order lower in 
$u$ than the pure Schwarzschild scenario. This implies the possibility of having a more pronounced 
stringy effect than the pure Einstein gravity. The dependence of the bending angle as well as the 
precession angle with a dilaton vacuum value has also been determined for different values of mass and charge. 
For the precession angle the charge dependence turns out to be opposite to that in R-N scenario, 
indicating that the dilaton coupled charged spacetime solution belongs to a new class of solutions other than the 
ordinary R-N scenario. Next in our analysis we have explored the role of the Gauss-Bonnet term at higher order. 
The modification of the perihelion shift due to this additional term puts a stringent bound on the 
parameter of the GB coupling. The Cassini effect as well as the Damour-Polyakov mechanism have been 
discussed in this context. Finally, we have considered quadratic gravity in a spherically symmetric spacetime. 
Here also a bound on the parameter of the model has been established after a critical analysis. 
As all these models are inspired by string theory, our work thus establishes bounds on stringy parameters 
in respect to present astrophysical observations.

\vskip 2cm
\noindent
Acknowledgement

S.C is funded by KVPY fellowship from DST, Government of India. He also thanks 
prof. Subenoy Chakraborty of Jadavpur University for helpful
discussions.

\end{document}